\documentclass[11pt,a4paper]{article}
\usepackage{jheparxiv}
\usepackage[utf8x]{inputenc}
\usepackage{amsmath}
\usepackage{amsfonts}
\usepackage{amssymb}
\usepackage{latexsym}
\usepackage{mathrsfs}
\usepackage{braket}		
\usepackage{color}
\usepackage{xcolor}
\usepackage{slashed}
\usepackage{twistor}
\usepackage[all]{xy}
\usepackage{tikz-cd}
\usepackage{mathtools}
\usepackage{bm}
\usepackage{color}
\usepackage{braket}
\usepackage{float}
\usepackage[]{todonotes}

\makeatletter
\renewcommand{\todo}[2][]{%
    \@todo[caption={#2}, #1]{\begin{spacing}{0.5}#2\end{spacing}}%
} 
\makeatother 
\usepackage{setspace}

\title{Cosmology and the classical limit of the S-matrix}
\author[a]{Katsuki Aoki}
\affiliation[a]{Center for Gravitational Physics and Quantum Information,
Yukawa Institute for Theoretical Physics, Kyoto University, 606-8502, Kyoto, Japan}

\author[a,b]{ and Andrea Cristofoli\,} 
\affiliation[b]{School of Mathematics and Maxwell Institute for Mathematical Sciences University of Edinburgh, EH9 3FD, United Kingdom}

\abstract{We investigate the relationships between classical observables in cosmology and the classical limit of quantum scattering amplitudes. We first look at the relation between Bogoliubov transformations and the notion of classical limit. Then, we compute the cosmological redshift for a particle in a cosmological background and the emitted gravitational waveform from a quantum field theory basis and its classical limit. We observe that there is no interpretation for the geodesic redshift purely in terms of on-shell amplitudes in flat space, given that the classical limit of a scalar 2-point vanishes when considering an FRW background with two asymptotically flat in and out regions. We resolve this apparent conundrum and recover the correct observable by showing that the action of Hermitian operators differs between the in and out regions, unlike standard approaches in flat spacetime. We then show that radiation reaction corrections to the redshift enter already at order $G$. Furthermore, we demonstrate that the emitted waveform can be represented solely in terms of an on-shell $3$-point amplitude in flat space without energy conservation, providing a closed formula for the waveform in an impulsive FRW. }

\begin{document}
{\baselineskip0pt
\rightline{\baselineskip16pt\rm\vbox to-20pt{
           \hbox{YITP-24-16}
\vss}}%
}

\maketitle

\section{Introduction}

Recent years have seen a surge of interest in the application of high-energy physics tools to reformulate perturbation theory in classical physics \cite{Buonanno:2022pgc}. Despite the main motivation, lying in the request for high precision in ongoing gravitational wave experiments, this programme is also testing a fundamental assumption about quantum field theory (QFT). Namely, every phenomenon --- either classical or quantum --- can be described in terms of on-shell scattering amplitudes in flat space \cite{Dixon:2011xs}. Applications in this context have already shown great promise. In the post-Minkoswkian approximation to the gravitational two-body problem \cite{Bjerrum-Bohr:2022blt}, state-of-the-art results in general relativity were derived using only on-shell amplitudes, without ever using a geodesic equation or the Einstein equations \cite{Bern:2019crd, Bern:2019nnu}. In a cosmological context, cosmological correlators on the future spacelike surface have singularities whose residue is a flat on-shell scattering amplitude \cite{Maldacena:2002vr,Maldacena:2011nz,Arkani-Hamed:2015bza}.
Even though these are unphysical singularities, in the sense that no real momenta can actually probe them, their existence and structure actually control the form of physical cosmological observables thus providing a beautiful connection between the study of the wavefunction of the universe and that of the S-matrix~\cite{Benincasa:2022gtd, Baumann:2022jpr}. It is then natural to explore the intersection of these two programmes and ask whether we can describe perturbatively the classical dynamics, in other words, timelike correlations, in an expanding and contracting universe from the classical limit of on-shell amplitudes in flat space. 

A natural place to address this question is to consider the so-called KMOC (Kosower-Maybe-O'Connell) formalism \cite{Kosower:2018adc} which naturally translates the classical limit of on-shell data into measurable physical quantities. Nowadays, it represents a robust framework out of which several phenomena relevant to classical general relativity can be studied such as the emission of gravitational waveforms, the bending of light, and the scattering of point particles with multipole and finite size effects \cite{Kosower:2022yvp}. However, despite its success, little is known about its feasibility to include cosmological effects. In order to understand whether a purely flat spacetime formulation capable of encoding cosmological effects is possible, and if not, what ingredients are needed, we will first consider the formulation of the KMOC formalism on a curved background \cite{Adamo:2022rmp}. The advantage of this method is twofold: on the one hand, it allows us to define the evolution of initial states using an S-matrix, with the interpretation of dynamical correlators in terms of flat space amplitudes deferred to a subsequent stage; at the same time, it offers an interesting framework where we can easily observe several assumptions in \cite{Kosower:2018adc} no longer holding. For instance, using this strategy it can be shown that the KMOC formalism on a plane wave background, is equivalent to its flat spacetime counterpart only if large gauge transformations are taken into account in the LSZ prescription in flat space \cite{Cristofoli:2022phh}. 

In this paper, we consider the KMOC formalism on a Friedmann–Robertson–Walker (FRW) background with two asymptotically flat regions at early and late times as a toy model. As is well known, several subtleties arise when doing QFT on a curved background, such as the definition of proper observables, the non-uniqueness of the vacuum, and the ambiguous notion of particles \cite{Birrell:1982ix,Hollands:2014eia, Parker:2009uva}. The FRW background having two asymptotically flat regions is one of the simplest spacetimes to address these issues and their interplay with the notion of classical limit. While still having an S-matrix and incoming and outgoing scattering regions, we will be able to trace relations with classical physics in the FRW background and --- when available --- with perturbation theory in flat space. 

The plan of the paper is as follows. In section 2, we reconsider the notion of the initial semiclassical state with the aim of representing point particles in a free incoming region of a cosmological background. Within this framework, the Bogoliubov transformation for massive modes results in the identity near the classical limit, effectively addressing the well-known ambiguity in the notion of point particles when approaching the classical limit in QFT on a curved background. However, we will observe that this does not hold true for massless modes and another treatment is needed for massless particles. Moving on to Section 3, we will describe the main elements defining the dynamical evolution of such states by the S-matrix. Here, we will calculate two processes in the perturbiner method: a massive scalar $2$-point amplitude and a $3$-point amplitude for a massive scalar emitting a graviton on an FRW background. Notably, we will find that the first quantity consistently vanishes, while the interpretation of the $3$-point process suggests equivalence to a $3$-point amplitude in flat space, albeit with the loss of energy conservation. In the subsequent Section 4, we further elucidate the definition of observable in this cosmological context showing its application for two classical observables: the cosmological redshift experienced by a massive point particle, as well as its emission of gravitational waves while moving through an FRW spacetime. We will then conclude by highlighting future directions and open questions. In Appendix~\ref{sec:classical}, we compute the gravitational wave emissions based on the classical equations of motion and confirm the agreement with the S-matrix approach.

\newpage

\section{The KMOC formalism on FRW backgrounds}\label{sec:KMOC}

The KMOC formalism is an on-shell approach for computing classical observable quantities starting from a first quantum field theory basis \cite{Kosower:2018adc}. Recently, it has been the object of major attention in the amplitude community, where several state-of-the-art results have been provided for the two-body problem in general relativity in the so-called post-Minkowskian approximation, which entails a weak field expansion around a flat spacetime. While its extension to curved backgrounds, such as plane wave backgrounds, has been investigated in \cite{Adamo:2022rmp} and \cite{Cristofoli:2022phh}, so far the KMOC formalism had little to say about the inclusion of cosmological effects into observables. To address this point, we will start by revisiting several aspects of QFT in cosmological backgrounds.
We work in the mostly negative signature and use hats on integral measures and delta functions to denote factors of $2\pi$ following the notation and normalization conventions set up in \cite{Kosower:2018adc}
\begin{equation}
\hat{\delta}^{(n)}(p) :=(2 \pi)^n \delta^{(n)}(p) \quad , \quad \hat{d}^np:=\frac{d^np}{(2\pi)^n} \quad , 
\end{equation}
Unless specified otherwise, we work in natural units with $c=\hbar=1$ and the gravitational coupling will be denoted as $\kappa:=\sqrt{32 \pi G}$ where $G$ is Newton's constant. 

\begin{figure}[t]
    \centering
    \includegraphics[width=0.5\linewidth]{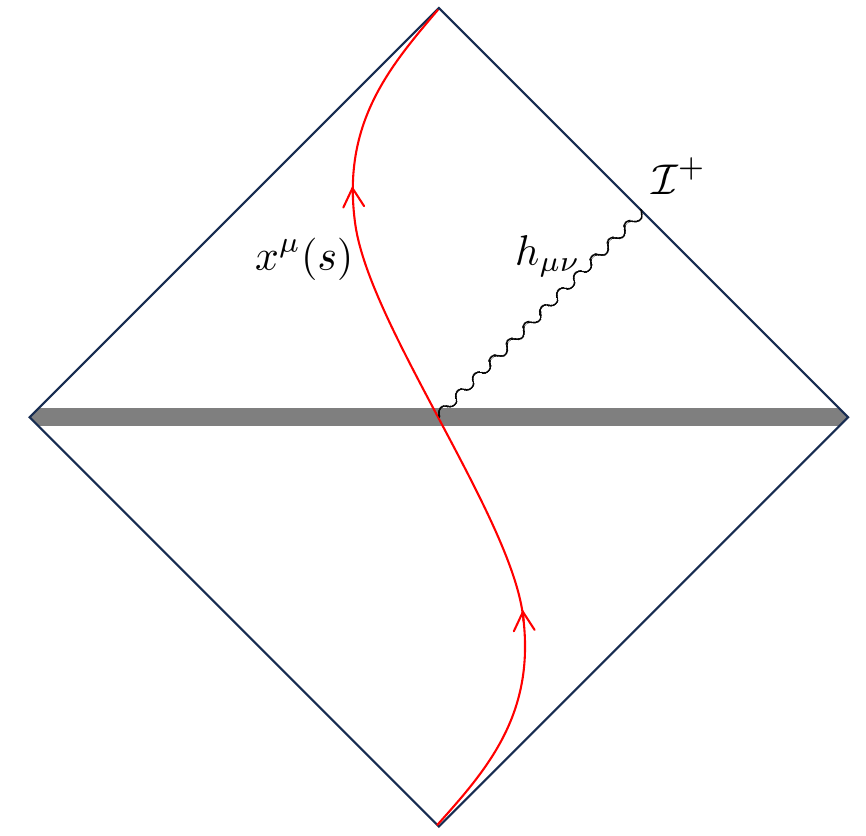}
    \caption{Penrose diagram with an FRW region sandwiched between two flat Minkowski regions. We will study a quantum description of the geodesic motion of a massive particle (red curve) and the emission of gravitational waves during passing in the FRW region (wavy line).}
    \label{fig:FRW}
\end{figure}

We then consider, instead of a flat background as in \cite{Kosower:2018adc}, a cosmological FRW background corresponding to two flat Minkowski regions at early and late times (Fig.~\ref{fig:FRW})
\begin{equation}
\begin{gathered}\label{eq:line-element}
    ds^2=a^2(\eta) \big( d\eta^2-dx^{i} dx^{j} \delta_{ij}\big)  \\ \lim_{\eta \rightarrow -\infty}a(\eta)=1 \quad , \quad \lim_{\eta \rightarrow +\infty}a(\eta)=a_{\infty} \in \mathbb{R}^{+} \ .
    \end{gathered}
\end{equation}
The scale factor $a(\eta)$ is a solution to Friedmann equations for a perfect fluid with density $\rho$ and pressure $p$
\begin{equation}
a'(\eta)^2=\frac{8 \pi G\rho(\eta) }{3} a^4(\eta) \quad , \quad a''(\eta)=\frac{4\pi G}{3} (\rho(\eta)-3p(\eta))a^3(\eta) \ .
\end{equation}
We won't be concerned with the particular realization of such matter distribution and the energy condition of matter, but simply assume its existence. The study of this spatially flat and isotropically changing metric as a background was used for the first time by Parker to investigate particle production in a simple cosmological setting \cite{Parker:1969au,Parker:1971pt}, and we will simply employ the same approach. 

We consider now the second quantization of a Klein-Gordon field $\phi$ on our FRW background, whose quanta will be used to model a semiclassical state describing a point particle on an expanding and contracting spacetime. By working with conformal time, we can represent the action of the scalar field $\phi$ on an FRW background as that of an auxiliary free field $\chi$ in Minkowski space with a time-dependent mass
\begin{align}
S&:=
\frac{1}{2}\int \sqrt{-g}d^4x \Big(g^{\mu\nu}\partial_{\mu}\phi\partial_{\nu}\phi-m^2 \phi^2 \Big)
\nonumber \\
&\,=\frac{1}{2}\int d^3x d\eta \Big[ (\chi')^2-(\nabla \chi)^2-(m^2a^2-a''/a)\chi^2 \Big]\ , 
\end{align}
with $\chi:=a \phi$.
The quantization of this field can be carried out by considering the mode expansion of $\chi$. Using bold symbols to denote spatial three-vectors,
\begin{equation}\label{eq:chi-2nd-quant}
\hat{\chi}(\mathbf{x}, \eta)=\int d\Phi(p) \left(e^{i \mathbf{p} \cdot \mathbf{x}} u_p(\eta) \hat{a}_{\mathbf{p}}+e^{-i \mathbf{p} \cdot \mathbf{x}} u_p^*(\eta) \hat{a}_{\mathbf{p}}^{\dag}\right),
\quad
d\Phi(p):=\frac{\hat{d}^3 \mathbf{p}}{2E_p(-\infty)}
\end{equation}
where the mode functions $u_p(\eta)$ are obeying the equations
\begin{equation}\label{eq:mode-f}
u_p^{\prime \prime}+\mathcal{E}_p^2(\eta) u_p=0, \quad \mathcal{E}_p(\eta) := \sqrt{|\mathbf{p}|^2+m^2a^2-a''/a}     \ ,
\end{equation}
and $E_p(-\infty)=\sqrt{|\mathbf{p}|^2+m^2}$ is the early time value of $\mathcal{E}_p(\eta)$. The mode functions only depend on the modulus $|\mathbf{p}|$ because of the spatial homogeneity and isotropy.
The associated creation and annihilation operators, $\hat{a}_{\mathbf{p}}$ and $\hat{a}^{\dag}_{\mathbf{p}}$, will satisfy the standard commutation relations as long as the mode functions (\ref{eq:mode-f}) are chosen to have the following Wronskian $W(u,u^*):=u'u^{*}-u u^{*}{}'=-2iE_p(-\infty)$,
\begin{equation}\label{eq:crea-annihi}
\left[\hat{a}_{\mathbf{p}}, \hat{a}_{\mathbf{p}^{\prime}}^{\dag}\right]=2E_p(-\infty)\hat{\delta}^{(3)}\left(\mathbf{p}-\mathbf{p}^{\prime}\right), \quad\left[\hat{a}_{\mathbf{p}}, \hat{a}_{\mathbf{p}^{\prime}}\right]=[\hat{a}_{\mathbf{p}}^{\dag}, \hat{a}_{\mathbf{p}^{\prime}}^{\dag}]=0 .
\end{equation}

Let's pause now to appreciate the first subtlety of doing QFT on a cosmological background as opposed to a flat one. In the standard formulation of KMOC, we implicitly select positive solutions to the Klein-Gordon equation with the energy $E_p(-\infty)$ being positive. This choice allows the operator $\hat{a}_\bold{p}^\dagger$ to be interpreted as the creation operator for a positive-frequency mode travelling forward in time. However, this interpretation relies on the existence of a global timelike Killing vector, but in our cosmological background such a Killing vector doesn't exist. An important consequence of this is that we can choose different mode functions $(u_{p}(\eta),u_{p}(\eta)^*)$ for expanding the auxiliary field $\chi$, and to define a set of creation and annihilation operators. For example, we can choose mode functions $(u_{p}^{in}(\eta),u_{p}^{in}(\eta)^*)$ which behaves as plane waves in the incoming region of our FRW background; alternatively, we can impose the same condition in the outgoing region for modes $(u_{p}^{out}(\eta),u_{p}^{out}(\eta)^*)$: the two choices allow for the definition of two inequivalent sets of creation and annihilation operators namely ($\hat{a}^{in}_\bold{p},(\hat{a}_\bold{p}^{in})^{\dag}$) and ($\hat{a}^{out}_\bold{p},(\hat{a}_\bold{p}^{out})^{\dag}$). The relation between the two can be found by noticing that both sets of modes are a basis of solutions to (\ref{eq:mode-f}). Thus, there exists a linear transformation relating the two 
\begin{equation}
u_p^{out}(\eta)=\alpha_p^* u_p^{in}(\eta)-\beta_p u_p^{in}(\eta)^*,
\end{equation}
with $\eta$-independent complex coefficients $\alpha_p$ and $\beta_p$. From this, it follows that 
\begin{equation}
\hat{a}_{\mathbf{p}}^{out}=\alpha_p \hat{a}_{\mathbf{p}}^{in}+\beta_p^* (\hat{a}_{-\mathbf{p}}^{in})^{\dag}, \quad (\hat{a}_{\mathbf{p}}^{out})^{\dag}=\alpha_p^* (\hat{a}_{\mathbf{p}}^{in})^{\dag}+\beta_p \hat{a}^{in}_{-\mathbf{p}} .
\end{equation}
Such transformation, relating inequivalent sets of modes, is known as \emph{Bogoliubov transformation}, while $\alpha_p$ and $\beta_p$ are $\mathbb{C}$-valued coefficients referred to as \emph{Bogoliubov coefficients}. If both sets of modes are normalized to have the same Wronskian $W=-2iE_p(-\infty)$, we have
\begin{equation}\label{eq:normali}
\left|\alpha_p\right|^2-\left|\beta_p\right|^2=1 \ .
\end{equation}
In particular, these coefficients are independent of $\eta$, and they can be uniquely defined as a Klein-Gordon inner product on any Cauchy hypersurface
\begin{equation}\label{eq:KG-in-pro}
\alpha_p=\frac{(u_p^{in})' (u_p^{out})^*-u_p^{in} (u_p^{out})^{* \prime}}{-2iE_p(-\infty)}, \quad {\beta_p}=\frac{(u_p^{in})^{\prime} u_p^{out}-u_p^{in} (u_p^{out})^{\prime}}{2iE_p(-\infty)}
\,.
\end{equation}

Since each set of modes defines a distinct set of creation and annihilation operators, the notion of vacuum --- defined as the state annihilated by an annihilation operator for all $\bold{p}$ --- becomes non-unique as long as $\beta_p \neq 0$. In particular, the vacuum with respect to the $\hat{a}^{out}_{\mathbf{p}}$ operator, denoted as $\ket{0_{out}}$, is non-trivial and it can be represented as a squeezed coherent state \cite{Schumaker:1986tlu,Grishchuk:1989ss,Parker:1969au, Parker:1971pt} with respect to the vacuum defined by $\hat{a}^{in}_{\mathbf{p}}$ which we will denote as $\ket{0_{in}}$. In terms of the Bogoliubov coefficients $\alpha_{p}$ and $\beta_{p}$ 
\begin{equation}\label{eq:squeezed-states}
\ket{0_{out}}=\mathcal{N} \exp \left(-\int d\Phi(p) \frac{\beta_{p}^*}{2 \alpha_{p} }(\hat{a}_{\mathbf{p}}^{in})^{\dag} (\hat{a}^{in}_{-\mathbf{p}})^{\dag}\right)\ket{0_{in}} \ ,
\end{equation}
where $\mathcal{N}$ is a proper normalization constant ensuring that the squeezed state is normalized. This non-uniqueness of vacua underlies several fascinating aspects of QFT on a curved background, such as the Unruh effect \cite{Davies:1974th, Unruh:1976db} and pair production in cosmological backgrounds \cite{Parker:1969au, Parker:1971pt}. In the KMOC formalism, however, it poses a potential complication. On the one hand, the non-uniqueness of vacua makes the notion of the S-matrix ambiguous.
At the same time, we are confronted with the fact that the notion of a point particle in QFT on a curved background is ambiguous: how can we arrive at the notion of a classical point particle upon which every observer should agree? 

To answer this question, we will examine in more detail the interplay between Bogoliubov transformations and the notion of classical limit.
We start by considering the Bogoliubov coefficients for massive modes. 
By restoring $\hbar$ factors in (\ref{eq:mode-f}), one can see that the solution is given by a WKB form in the classical limit. More precisely, by restoring $\hbar$ factors, it can be noticed that $\mathcal{E}_p^2/\hbar^2=E_p^2/\hbar^2-a''/a$ where $E_p=\sqrt{|\mathbf{p}|^2+m^2a^2}=[\text{mass}]$ is the energy of classical point particle. On the other hand, the curvature is $[\text{lenght}]^{-2}$, so the curvature effects are negligible as $\hbar \rightarrow 0$. We can thus summarize the expression for the semiclassical states for a massive mode as
\begin{equation}\label{eq:WKB}
    u_p(\eta)=e^{-i \int_{-\infty}^{\eta}d\eta' \: E_{p}(\eta')/\hbar} \sqrt{\frac{E_{p}(-\infty)}{E_{p}(\eta)}}   \ ,
\end{equation}
under our normalization condition.
Within this approximation, the late-time behaviour of the WKB mode with incoming boundary conditions is equivalent to the same mode with outgoing conditions, eliminating the distinction between incoming and outgoing boundary conditions, namely $\alpha_{p}=1$ and $\beta_p=0$. This implies the equivalence of (adiabatic) vacua for massive modes in the in and out regions \cite{Parker:2009uva}. Most importantly, every observer will agree on the notion of a point particle in the semiclassical limit as in the standard formulation of the KMOC formalism on a flat background.

For massless modes, on the other hand, the situation is more subtle. As demonstrated in \cite{Kosower:2018adc}, in the classical limit for massless modes, what should be kept fixed is the wavenumber $\bar{k}=k/\hbar$, rather than the momenta $k$ of the mode. This implies that a WKB approximation to (\ref{eq:mode-f}) for massless modes cannot be applied in the classical limit. Consequently, the notion of vacua for massless particles becomes ambiguous, raising concerns about pair production in the classical limit even though there should be no pair production classically. We can check this by considering a concrete example of FRW background from \cite{Birrell:1982ix}
\begin{equation}\label{eq:line-element-impulsive}
a^2(\tau):=\bigg(\frac{a^4_{\infty}+1}{2}+\frac{a^4_{\infty}-1}{2} \tanh(\tau \gamma)\bigg)^{1/2} \quad , \quad \tau:=\int^{\eta}\frac{d\eta'}{a^2(\eta')} \ ,
\end{equation}
where $\gamma$ is a positive real number. The associated ratio of Bogoliubov coefficients, for massless modes with wavenumber $\bar{k}$, can be easily computed from the Klein-Gordon inner product (\ref{eq:KG-in-pro}). Restoring $\hbar$ factors and denoting the frequency as $\omega$, the Bogolivbov coefficients are computed as
\begin{align}
    \alpha_{\bar k} &=  \frac{a_{\infty} \: \Gamma(1-\frac{i\omega}{\gamma})\Gamma(-\frac{i\omega a^2_{\infty}}{\gamma})}{\Gamma\big(-\frac{i(1+a^2_{\infty})\omega}{2\gamma}\big)\Gamma \big(1-\frac{i\omega(1+a^2_{\infty})}{2\gamma}\big)}
    \,,\\
    \beta_{\bar k} &= \frac{a_{\infty} \:  \Gamma(1-\frac{i\omega}{\gamma})\Gamma(\frac{i\omega a^2_{\infty}}{\gamma})}{\Gamma\big(-\frac{i\omega(1-a^2_{\infty})}{2\gamma}\big)\Gamma \big(1-\frac{i\omega(1-a^2_{\infty})}{2\gamma}\big)}
    \label{eq:beta}
\end{align}
and then
\begin{equation}\label{eq:ratio-alfa-beta}
\left|\frac{\beta_{\bar k}}{\alpha_{\bar k}}\right|^2=\frac{\sinh ^2\left[\frac{\pi\left(1-a_{\infty}^2\right)\omega}{2 \gamma } \right]}{\sinh ^2\left[\frac{\pi\left(1+a_{\infty}^2\right)\omega}{2 \gamma} \right]}  \neq 0 \ .
\end{equation}
One could then worry that massless particle production seems to be relevant.

However, we can verify that such particle production only contributes with quantum effects to classical observables. Consider for example a set of modes for scalar massless particles (the same applies to gravitons). The on-shell number operator of outgoing modes is \cite{Cristofoli:2021vyo}
\begin{equation}
\mathbb{N}_{out}=
\int d \Phi(k) (\hat{a}_{\bold{k}}^{out})^{\dag} \hat{a}^{out}_{\bold{k}} \ .
\end{equation}
With respect to the vacuum $\ket{0_{in}}$, the mean density of massless particle is\footnote{The quantity $\hat{\delta}^{(3)}(0)$ arises from the standard commutation relation (\ref{eq:crea-annihi}). However, it is not physically relevant to this discussion and can be easily avoided by working on a finite Cauchy slice of the manifold.}
\begin{equation}\label{deltaO}
 \bra{0_{in}}\mathbb{N}_{out}\ket{0_{in}}=\hat{\delta}^{(3)}(0)\int \hat{d}^3\bold{k} |\beta_{k}|^2= \hat{\delta}^{(3)}(0)  \hbar^3 \int \hat{d}^3\bold{\bar{k}} |\beta_{\bar{k}}|^2 \sim_{\hbar \rightarrow 0} 0 \ ,
\end{equation}
where we note that $\beta_{\bar k}$ \eqref{eq:beta} is finite in the classical limit.
Thus, the vacuum state $\ket{0_{in}}$ contains no additional particles with respect to $\ket{0_{out}}$ implying that quantities such as the emitted power and waveform, computed with respect to the vacua $\ket{0_{in}}$ or $\ket{0_{out}}$, 
won't be affected by particle production. 
Although this argument is based on a specific choice of background, and thus of Bogoliubov coefficients \eqref{eq:beta}, we can see that this statement should be more general. The only way this argument would fail is if $\beta_{k}$ were to have a Laurent expansion in $1/\hbar$. However, if that were the case the Bogoliubov transformation for massless modes would become singular thus preventing to have well-defined scalar products in our Hilbert space.

There is yet another way to see the irrelevance of the massless particle production in the classical limit. In the classical context, what is of interest about a massless field is a classical {\it wave} described by that field. Such a wave is quantum-mechanically described by a coherent state $\ket{\alpha}$ with a large number of particles such that~\cite{Cristofoli:2021vyo}
\begin{align}
    \lim_{\hbar \to 0} \bra{\alpha} \hat{\chi} \ket{\alpha} = \chi_{\rm cl} \neq 0\,.
\end{align}
On the other hand, one can easily show
\begin{align}
    \bra{0_{in}}\hat{\chi} \ket{0_{in}} = \bra{0_{out}}\hat{\chi} \ket{0_{out}} = 0
    \,,
\end{align}
i.e., both $\ket{0_{in}}$ and $\ket{0_{out}}$ do not observe any classical waves. Although $\ket{0_{in}}$ and $\ket{0_{out}}$ are inequivalent, both can be equally regarded as a classical vacuum of waves.

To summarize, in the KMOC formalism on a cosmological background, the non-equivalence of vacua does not pose an obstacle to the definition of on-shell formulation of observables. We can safely focus on the change of the in-in expectation value between the in region at $\eta=-\infty$ and in the out region at $\eta=+\infty$
\begin{align}
    \Delta O = \langle \Psi|\mathcal{S}^{\dagger}\mathcal{O}(+\infty) \mathcal{S}|\Psi\rangle - \langle \Psi|\mathcal{O}(-\infty)|\Psi\rangle \,,
\end{align}
Here, $\ket{\Psi}$ is a properly defined initial state, analogous to that in flat space. $\mathcal{S}$ represents the time evolution operator, which will be the primary focus in the next section, while $\mathcal{O}(\eta)$ denotes a composite operator of the fields, such as a momentum operator or a waveform operator. The main difference in formulating these operators on a curved manifold lies in their dependence on the location of the manifold. Indeed, on the FRW background, Hermitian operators $\mathcal{O}(\eta)$ generally exhibit time dependence. This dependence is crucial for accurately reproducing classical observables in cosmological backgrounds, as we will see in Sec.~\ref{sec:observables}.

\section{Scattering amplitudes on FRW backgrounds}

While particle production can be ignored in the classical limit, we should first consistently use states and operators in the intermediate quantum-based computations, especially for massless particles, and then take the classical limit. It would be intuitive to describe the initial state $\ket{\Psi}$ by using the in vacuum so we decide to use them. While the KMOC formalism computes the in-in expectation, its building blocks are in-out correlations, namely scattering amplitudes. In curved spacetime, a natural scattering amplitude would be the amplitude between the in/out states defined by the in/out vacua. On the other hand, our observable is the in-in expectation so the operators are always sandwiched by the free states defined by the in vacuum (we shall use the completeness relation spanned by the in-particle states when it is inserted). Therefore, the building blocks of the KMOC formalism are the scattering amplitudes where both in/out states are defined by the same definition of the positive frequency mode, say
\begin{align}
    \bra{p_2}\mathcal{S}\ket{p_1}\,, \quad {\rm with} \quad \ket{p_1}=(\hat{a}_{\mathbf{p}_1}^{in})^{\dagger}\ket{0_{in}}\,, \quad
    \bra{p_2}=\bra{0_{in}}\hat{a}_{\mathbf{p}_2}^{in}
    \,.
\end{align}
Hereinafter, we shall omit in/out labels because the vacuum is always defined by the in region.

We evaluate such scattering amplitudes on recursively constructed solutions to the equations of motion using the perturbiner approach \cite{Arefeva:1974jv,Jevicki:1987ax,Mizera:2018jbh, Selivanov:1997aq, Cho:2023kux}. The on-shell action plays the role of a generating function of the tree-level S-matrix. One of the advantages of this approach is that it replaces the combinatorial computations of Feynman rules with the study of differential equations with proper boundary conditions. These solutions have boundary conditions corresponding to asymptotic scattering states, aligning with an LSZ truncation, and are often nontrivial. When the background fields allow for an S-matrix, such as in shock waves and sandwich plane waves backgrounds in gravity, the perturbiner approach not only reproduces the amplitudes derived using standard Feynman rule computations on a background but also resums an infinite number of Feynman diagrams in flat space. On cosmological backgrounds, such as an asymptotically de Sitter universe, the perturbiner approach has been already employed to compute relevant observables as the so-called wavefunction of the universe related to quantum mechanical observables \cite{Albayrak:2023hie,Benincasa:2022gtd}. In this approach, these observables are considered directly as functions of the data at the space-like boundary at future infinity, disregarding explicit time evolution. Instead, we consider a cosmological background that returns asymptotically flat after a given amount of time such as (\ref{eq:line-element}) and discuss the time evolution from the past flat region to the future flat region, which is analogous to calculations on a plane wave background~\cite{Adamo:2017nia}. Let's then proceed with the detailed evaluation of the amplitudes. We will focus on two processes: a scalar $2$-point amplitude for a massive particle on FRW, and a $3$-point for a massive scalar particle emitting a graviton on the background. As we will see, the first process is related to the cosmological redshift while the second one is to the emission of a gravitational waveform for a point particle crossing the background.

\subsection{Vanishing 2-point amplitudes in FRW}
\label{sec:2pt}
Consider a scalar $2$-point amplitude for a massive particle on FRW. Following the perturbiner approach, we can compute this process as the quadratic part of a massive Klein-Gordon action minimally coupled to (\ref{eq:line-element}), and evaluate it on the respective solutions to the equations of motion. Defining the $S$-matrix on the background as $\mathcal{S}=1+i \mathcal{T}$, the perturbiner gives
\begin{align}
    i \: \bra{p_2}\mathcal{T}\ket{p_1} & = {\frac{\delta^2}{\delta \epsilon_1 \delta \epsilon_2} S|_{\phi=\epsilon_1 \phi_1 +\epsilon_2 \phi_2^*, g_{\mu\nu}=\bar{g}_{\mu\nu}}
    }
    \nonumber \\
    &=\int_{\mathbb{R}^4}d^{4}x \: a^4(\eta) \bigg(\bar{g}^{\mu \nu}(x)\partial_{\mu}\phi_{2}^{*}(x)\partial_{\nu}\phi_{1}(x)-m^2\phi_{2}^{*}(x)\phi_{1}(x) \bigg) \: ,
\end{align}
at tree level where $\bar{g}_{\mu\nu}=(a^2,-a^2 \delta_{ij})$ is the background metric. For $\phi_1$ and $\phi_2$, we are considering the solutions to the equations of motions under the positive frequency boundary condition at the past infinity $\eta \to -\infty$:
\begin{align}\label{eq:boundary-BD}
    \phi_{A}(-\infty) &\to e^{-ip_{A\mu} x^{\mu}} \quad , \quad p_{A\mu}=(\sqrt{|\mathbf{p}_A|^2+m^2},-\mathbf{p}_A)
    \,, \quad A=1,2
    \,.
\end{align}
As for the structure of the $2$-point amplitude, the homogeneity and isotropy of the background are reflected in the conservation of spatial momenta in a scattering process. This is easily noticed by integrating out the spatial components of the wave solutions
\begin{align}
i \: \bra{p_2}\mathcal{T}\ket{p_1} &= \hat{\delta}^{(3)}(\mathbf{p}_2-\mathbf{p}_1)\int_{-\infty}^{+\infty} d\eta \: \bigg( u_2^{*\,'}(\eta)u_1'(\eta)-\mathcal{E}^2_1(\eta) u_2^{*}(\eta) u_1(\eta)\bigg) \nonumber \\
&= \hat{\delta}^{(3)}(\mathbf{p}_2-\mathbf{p}_1)\int_{-\infty}^{+\infty} d\eta \: \bigg( u_1^{*\,'}(\eta)u_1'(\eta)-\mathcal{E}^2_1(\eta) u_1^{*}(\eta) u_1(\eta)\bigg) \ .
\end{align}
Here, the mode function $u_A$ is introduce in the same way as Sec.~\ref{sec:KMOC} and we use the shorthand $\mathcal{E}^2_A=\mathcal{E}^2_{p_A}$.
By rewriting the integrand via integration by parts, we can see that the required $2$-point amplitude is the sum of two terms\footnote{Such statement is general to $2$-point amplitudes in background and it simply follows from the possibility to rewrite the action via Stokes theorem.}
\begin{align}
\label{eq:boundary-FRW}
i \: \bra{p_2}\mathcal{T}\ket{p_1}&= \hat{\delta}^{(3)}(\mathbf{p}_2-\mathbf{p}_1) u_1^{*}(\eta) u_1'(\eta)|^{\eta=+\infty}_{\eta=-\infty}  \nonumber \\
&-\hat{\delta}^{(3)}(\mathbf{p}_2-\mathbf{p}_1)\int_{-\infty}^{+\infty} d\eta \: u_{1}^{*}(\eta) \bigg( u_1''(\eta)+\mathcal{E}^2_1(\eta)u_1(\eta)\bigg) \ .
\end{align}
The second term is zero on-shell as a consequence of (\ref{eq:mode-f}). Thus, our 2-point amplitude can be interpreted as a pure boundary term coming from the two spatial Cauchy slices at $\eta=+\infty$ and $\eta=-\infty$
\begin{equation}\label{eq:result-2-point}
i \: \bra{p_2}\mathcal{T}\ket{p_1}= \hat{\delta}^{(3)}(\mathbf{p}_2-\mathbf{p}_1) u_1^{*}(\eta) u_1'(\eta)|^{\eta=+\infty}_{\eta=-\infty} \ .
\end{equation}
We use the WKB approximation (\ref{eq:WKB}) in looking at the leading classical limit of our $2$-point amplitude. Then, we derive the result that the $2$-point amplitude is vanishing on the FRW background 
\begin{equation}\label{eq:2-point-FRW-vanishing}
   i    \bra{p_2}\mathcal{T}\ket{p_1} =\hat{\delta}^{(3)}(\mathbf{p}_2-\mathbf{p}_1) \bigg( u_1^{*}(+\infty) u_1'(+\infty)-u_1^{*}(-\infty) u_1'(-\infty) \bigg)=0
   \,.
\end{equation}
At first glance, this result may seem odd in light of the fact that tree-level $2$-point amplitudes on backgrounds should encode the geodesic motion for a point particle in the related spacetime. While this is certainly true for a large class of solutions to the Einstein equations, such as Schwarzschild \cite{Kol:2021jjc}, plane waves and gravitational shock waves \cite{Adamo:2022rmp}, it is no longer valid for the case of FRW. This is the first example of the subtlety between the formulation of KMOC on flat spacetime versus cosmological ones. In the latter, as shown in \eqref{deltaO}, a Hermitian operator for a given observable can be different in the in and out regions, thus implying that not every observable can be derived from scattering data. We will delve into more detail on this when discussing the extraction of classical observables.

\subsection{3-point amplitudes in FRW and Minkowski}
\label{sec:3pt}
We now turn our attention to a less trivial amplitude describing an incoming and outgoing scalar point particle emitting a graviton with helicity $\sigma$ on a cosmological background. Similar to the $2$-point amplitude, the tree-level $3$-point amplitude is computed by the $3$-linear part of the action perturbed around (\ref{eq:line-element}) and evaluated on solutions to the field equations of motion. Neglecting back-reaction with the background, we construct such amplitude by choosing the proper boundary conditions for the graviton
\begin{align}
    i\bra{p_2,k^{\sigma}}\mathcal{T}\ket{p_1}&=
    {\frac{\delta^3}{\delta \epsilon_1 \delta \epsilon_2 \delta \epsilon_3}S|_{\phi=\epsilon_1 \phi_1+\epsilon_2 \phi_2^*, g_{\mu\nu}=\bar{g}_{\mu\nu}+\epsilon_3 \kappa (h^{\sigma}_{\mu\nu})^*} }
    \nonumber \\
    &=- \kappa \int_{\mathbb{R}^4} d^{4}x \: a^4(\eta) (h^{\mu \nu}_{\sigma}(x))^*  \partial_{\mu}\phi_2^{*}(x)\partial_{\nu}\phi_1(x) \ ,
\end{align}
where $h^{\mu \nu}_{\sigma}(x)$ is a solution to linearized Einstein equations on a FRW background with incoming boundary condition, while $\phi_A(x)~(A=1,2)$ is defined as in the 2-point case. The contribution from the trace of $h^{\mu \nu}_{\sigma}(x)$ has been ignored, as the graviton corresponds to the transverse-traceless (TT) mode. The action can be rewritten solely in terms of flat spatial contractions
\begin{equation}
    i\bra{p_2,k^{\sigma}}\mathcal{T}\ket{p_1}=-\kappa \int_{\mathcal{\mathbb{R}}^3} d^{3}x \int_{-\infty}^{+\infty}d\eta \:  a^4(\eta) (h^{i j}_{\sigma}(\mathbf{x},\eta))^*  \partial_{i}\phi_2^{*}(\mathbf{x},\eta)\partial_{j}\phi_1(\mathbf{x},\eta) \ .
\end{equation}
Once again, homogeneity and isotropy of the background make our life easier so that we can make the following forms for the spatial behaviour of the states under consideration
\begin{equation}
    h^{i j}_{\sigma}(\mathbf{x},\eta):=\frac{\Psi(\eta)}{a^2(\eta)}  \varepsilon^{ij}_{-\sigma}(\mathbf{k}) e^{i \mathbf{k} \cdot \mathbf{x}} \quad , \quad \phi_A(\mathbf{x},\eta):=\frac{u_A(\eta)}{a(\eta)} e^{i \mathbf{p}_A \cdot \mathbf{x}} \ .
\end{equation}
Here, $\varepsilon^{ij}_{\sigma}$ is the polarization tensor same as the flat spacetime. The Einstein equation reads (see e.g.~\cite{Maggiore:2018sht})
\begin{equation}\label{eq:chi-def}
    \Psi''(\eta)+2\frac{a'(\eta)}{a(\eta)}{\Psi'(\eta)}+|\mathbf{k}|^2\: \Psi(\eta)=0 \ ,
\end{equation}
which we solve under the boundary condition $\Psi \to e^{-i|\mathbf{k}|\eta}$ as $\eta \to -\infty$.

We now have all the ingredients to write down a closed expression for a $3$-point amplitude on an FRW background in the absence of back-reaction
\begin{equation}
\label{eq:3-point}
    i\bra{p_2,k^{\sigma}}\mathcal{T}\ket{p_1}= -\kappa \: \hat{\delta}^{(3)}(\mathbf{p}_1-\mathbf{p}_2-\mathbf{k}) \varepsilon^{ij}_{\sigma}(\mathbf{k}) [\mathbf{p}_1 \mathbf{p}_2]^{TT}_{ij}  \int_{-\infty}^{+\infty} d\eta \: \Psi^*(\eta)u_2^{*}(\eta)u_1(\eta) \ ,
\end{equation}
where $[\mathbf{p}_1 \mathbf{p}_2]^{TT}_{ij}$ is the TT projection of $\mathbf{p}_1\mathbf{p}_2$ with respect to $\mathbf{k}$.
The spatial part has the same structure as a 3-point amplitude in a vacuum for a massive scalar particle emitting a graviton with helicity $\sigma$. Normally, the 3-point amplitude would vanish under real, on-shell kinematics and the conservations of energy and momentum. However, due to the absence of a Dirac delta function representing the conservation of energy (reflecting the non-conservation of energy in a FRW background), this 3-point amplitude does not vanish. Instead, this distribution has now been replaced by a non-vanishing integral over a conformal time. Notice that if the scale factor were equal to one for all times (i.e. a Minkowski background), then this integral would exactly become a Dirac delta, ensuring the conservation of energy, as expected and leading to a vanishing $3$-point. From this perspective, we might interpret the remaining integral as a deformation of the Dirac delta for the conservation of energies influenced by the presence of cosmological expansions and contractions. All in all, the 3-point amplitude in the FRW background can be understood as the on-shell 3-point in the flat space without energy conservation.

Let's study now the leading classical limit of (\ref{eq:3-point}). It is useful to reintroduce the dependence of all our quantities on $\hbar$, considering that the momenta of gravitons are quantum $k=\hbar \bar{k}$ in contrast to the momenta of massive particles. Here, $\bar{k}$ represents the wavenumber to be kept fixed while taking the classical limit. Another simplification within this limit arises from the observation that the classical limit of our $3$-point process is described by solutions in the WKB limit \eqref{eq:WKB}. Using these solutions as in the $2$-point case, we can represent the leading classical limit of our $3$-point amplitude as follows
\begin{equation}
\begin{gathered}
    i\bra{p_2,k^{\sigma}}\mathcal{T}\ket{p_1}= - \frac{\kappa}{\hbar^{3/2}}  \: \hat{\delta}^{(3)}(\mathbf{p}_1-\mathbf{p}_2-\hbar\bar{\mathbf{k}}) \varepsilon^{ij}_{\sigma}(\mathbf{k}) [\mathbf{p}_1 \mathbf{p}_2]^{TT}_{ij}   \\ 
    \times  \int_{-\infty}^{+\infty}d\eta \: \Psi^*(\eta)  \frac{e^{-i \int_{-\infty}^{\eta}d\eta' \: E_1(\eta')/\hbar}}{\sqrt{E_1(\eta)/E_1(-\infty)}} \frac{e^{i \int_{-\infty}^{\eta}d\eta' \: E_2(\eta')/\hbar }}{\sqrt{E_2(\eta)/E_2(-\infty)}} \ .
    \end{gathered}
\end{equation}
where one factor $\hbar^{1/2}$ comes from $\kappa$ and another $\hbar^{-1}$ from the measure of integration.
Notice that the interference between the WKB phases of the incoming and outgoing scalar particles reproduces a non-vanishing exponential term involving the wavenumber of the graviton, denoted as $\bar{k}$, as $\hbar \rightarrow 0$. As a result, we obtain
\begin{equation}
\begin{gathered}
    i\bra{p_2,k^{\sigma}}\mathcal{T}\ket{p_1}= -\frac{\kappa}{ \hbar^{3/2}} E_1(-\infty)\: \hat{\delta}^{(3)}(\mathbf{p}_1-\mathbf{p}_2-\hbar\bar{\mathbf{k}}) \varepsilon^{ij}_{\sigma}(\mathbf{k}) [\mathbf{p}_1 \mathbf{p}_2]^{TT}_{ij} \\ \times 
     \int_{-\infty}^{+\infty} d\eta\, \Psi^*(\eta) \frac{e^{-i \int_{-\infty}^{\eta}d\eta' \frac{\bold{p}_1 \cdot \bar{\bold{k}}}{E_1(\eta')}}}{E_1(\eta)} \ .
     \label{3pt_classical}
     \end{gathered}
\end{equation}
We have kept the $\mathbf{k}=\hbar \bar{\mathbf{k}}$ contribution in the Dirac delta since the ``momentum mismatch'' $\mathbf{p}_1-\mathbf{p}_2$ can scale as $\hbar$. As we will show in Section 4 this is crucial to reproduce radiation reaction corrections to the redshift. Having understood the $\hbar$ scaling and the classical limit, let us come back to the natural unit $\hbar =1$ for notational simplicity.

Eq.~\eqref{3pt_classical} is the closest expression we can provide for this process. In general, $\Psi(\eta)$ has to be found for a given choice of FRW dynamics, which can be a complicated task considering that for these modes, we cannot apply a WKB approximation\footnote{In contrast to the massive scalar modes, the ordinary differential equation defining the massless modes has no $\hbar$ factor associated with it. Consequently, we cannot ignore curvature effects on its dynamics.}. While, in general, it is not always possible to provide a closed analytic formula for such wave solutions, it is interesting to note that there exist choices of the scale factor that can provide simple results. To this end, we consider the scale factor previously discussed (\ref{eq:line-element-impulsive}). 
From \cite{Parker:2009uva}, a solution to the equations of motion can be given in a closed form in terms of a hypergeometric function
\begin{equation}
 \Psi(\tau)=e^{-i|\bold{\bar{k}}|\tau}   { }_2 F_1(c_2-c_1,-c_1-c_2;1-2c_1;-e^{2\tau \gamma}) \ ,
\end{equation}
where we have defined
\begin{equation}
    c_{1}:=\frac{i|\bold{\bar{k}}|}{2\gamma} \quad , \quad c_2:=\frac{i|\bold{\bar{k}}|a^2_{\infty}}{2\gamma} \ .
\end{equation}
From this, the $3$-point amplitude is
\begin{equation}
\begin{gathered}
    i\bra{p_2,k^{\sigma}}\mathcal{T}\ket{p_1}= -\kappa E_1(-\infty)\: \hat{\delta}^{(3)}(\mathbf{p}_1-\mathbf{p}_2-\mathbf{k}) \varepsilon^{ij}_{\sigma}(\mathbf{k}) [\mathbf{p}_1 \mathbf{p}_2]^{TT}_{ij} \\ 
    \times \int_{-\infty}^{+\infty}d\tau a^2(\tau) \:{ }_2 F_1^*(c_2-c_1,-c_1-c_2;1-2c_1;-e^{2\tau \gamma}) \frac{e^{i|\bold{\bar{k}}|\tau-i \int_{-\infty}^{\eta(\tau)}d\eta' \frac{\bold{p}_1\cdot \bar{\bold{k}}}{E_1(\eta')}}}{E_1(\eta(\tau))} \ .
    \label{3pt_tau}
    \end{gathered}
\end{equation}
In analogy with similar calculations on plane wave backgrounds, it is interesting to consider an impulsive limit of FRW by letting the parameter $\gamma$ go to infinity $\gamma \rightarrow +\infty$. Within this limit, the scale factor (\ref{eq:line-element-impulsive}) approaches a step function $a(\tau)=1+(a_{\infty}-1)\theta(\tau)$ and then
\begin{align}
\:{ }_2 F_1(c_2-c_1,-c_1-c_2;1-2c_1;-e^{2\tau \gamma})= 
    \begin{cases}
    1 & (\tau<0)\,, \\
    \alpha e^{i(1-a_{\infty}^2)|\bold{\bar{k}}|\tau} + \beta e^{i(1+a_{\infty}^2)|\bold{\bar{k}}|\tau} &(\tau>0)\,,
    \end{cases}
\end{align}
where the Bogoliubov coefficients for the massless modes are
\begin{equation}
\alpha=\frac{a_{\infty}^2+1}{2a_{\infty}}  \quad , \quad  \beta=\frac{a_{\infty}^2-1}{2a_{\infty}} \ .
\end{equation}
The integral of \eqref{3pt_tau} can be performed analytically by splitting the integration region into two subsets and with a proper $i \epsilon$ prescription. The final result is
\begin{equation}\label{fin:3-point}
\begin{gathered}
     \lim_{\gamma \rightarrow +\infty} i\bra{p_2,k^{\sigma}}\mathcal{T}\ket{p_1} = i \kappa E_1(-\infty)\: \hat{\delta}^{(3)}(\mathbf{p}_1-\mathbf{p}_2-\mathbf{k}) \varepsilon^{ij}_{\sigma}(\mathbf{k}) [\mathbf{p}_1 \mathbf{p}_2]^{TT}_{ij} \\ 
    \times 
    \bigg(\frac{1}{E_1(-\infty)|\bar{\bold{k}}|-\bold{p}_1 \cdot \bar{\bold{k}}-i \varepsilon}-\frac{\alpha}{E_1(+\infty)|\bold{\bar{k}}|-\bold{p_1} \cdot \bold{\bar{k}}+i\varepsilon} + \frac{\beta}{E_1(+\infty)|\bold{\bar{k}}|+\bold{p}_1 \cdot \bold{\bar{k}}-i\varepsilon}\bigg)
    \end{gathered}
\end{equation}
\newline
The last line in (\ref{fin:3-point}) provides an analytic representation of the modification of the conservation of energy in the chosen background. By using the Sokhotski-Plemelj theorem $\lim _{\varepsilon \rightarrow 0^{+}} \frac{1}{x \pm i \varepsilon}=\mp i \pi \delta(x)+\mathcal{P}\left(\frac{1}{x}\right)$ in the absence of cosmological effects~$(a_{\infty}=1)$ where $\alpha=1$ and $\beta=0$, it can then be easily seen that the first two terms add up to a Dirac delta for the conservation of energy, resulting in a vanishing three-point amplitude. More generally, though, this is non-zero with important consequences on classical cosmological observables. Having a modification of Dirac deltas when computing amplitudes on a background, is a standard feature of computing matrix elements of the S-matrix on a curved background. From the perspective of the perturbiner method, they simply reflect the fact that the chosen background has a different number of Killing vectors --- and thus of allowed isometries --- as opposed to the flat case. It is interesting to notice that such modifications can also be understood from a purely flat spacetime perspective, as coming from the insertion of large gauge transformations in the LSZ truncation. As shown in \cite{Cristofoli:2022phh}, these are related to memory effects and they are crucial to keep into account in order to reproduce the correct classical limit of observables in KMOC.

\section{Classical observables in cosmology from on-shell data}
\label{sec:observables}

We are now in a position to evaluate physical observables relevant in a cosmological context such as the redshift. We can start by considering a classical point particle with mass $m$, described by the following semiclassical initial state at early times
\begin{equation}\label{eq:initial-state}
    \ket{\Psi}:=\int d\Phi(p) \phi(p) \ket{p} \ .
\end{equation}
Here, $\phi(p)$ is a Lebesgue integrable function describing a wavepacket, as in \cite{Kosower:2018adc}. As for $\ket{p}$, it describes a massive particle state with momentum $p$ and mass $m$, constructed from the vacuum defined as the state minimizing the Hamiltonian operator at early times. Following the previous description of the Bogoliubov transformation in the classical limit, such a state is well-defined for all observers. In describing an observable like the redshift, all we need then is the $S$-matrix and a proper definition of the momentum operator. The $S$-matrix is provided by the scattering amplitudes computed in Section 3 via the perturbiner approach with the appropriate boundary conditions.
As for the momentum operator, this can be different at early and late times when working on a curved background, as opposed to the flat spacetime formulation of KMOC in \cite{Kosower:2018adc}. Let's look at this in more detail.
\newline
\subsection{The cosmological redshift and back-reaction}
We work on the previous example of FRW spacetime with coordinates $(\eta,\mathbf{x})$. We can then define a momentum operator for a massive particle with mass $m$ at time $\eta$ as
\begin{equation}
    \mathbb{P}_{\mu}(\eta):=\int d^3x \: T_{0 \mu} \:(\hat{\chi}(\mathbf{x}, \eta)\:,\:\partial \hat{\chi}(\mathbf{x}, \eta)) \ ,
\end{equation}
where $T_{0 \mu}$ is the energy-momentum tensor of a second quantized massive scalar field as in (\ref{eq:chi-2nd-quant}). 
Applying the WKB approximation \eqref{eq:WKB}\footnote{There is an implicit normal ordering in this expression. We do not emphasize it because our focus is on differences in the mean value of this operator, where the energy of the vacuum is subtracted.}, we obtain
\begin{equation}\label{eq:momentum-operator-FRW}
\mathbb{P}_{\mu}(\eta)=\int d\Phi(p) \: p_{\mu}(\eta)\: \hat{a}_{\mathbf{p}}^{\dag} \hat{a}_{\mathbf{p}} \quad , \quad p_{\mu}(\eta):=\bigg(\sqrt{|\mathbf{p}|^2+m^2a^2(\eta)},-\mathbf{p}\bigg) \ .
\end{equation}
Recall that the creation and annihilation operators are defined by the in mode function.
From (\ref{eq:momentum-operator-FRW}), now we can clearly see that the momentum operator $\mathbb{P}_{\mu}$ is not the same in the in and out region, defined respectively by $\eta \rightarrow -\infty$ and $\eta \rightarrow +\infty$, due to $p_{\mu}(\eta)$. Let's examine the impact of this time dependence on evaluating in KMOC. We define the impulse experienced by the particle by working with coordinates $(\eta, \mathbf{x})$ 
\begin{equation}\label{eq:impulse-redshift}
    \Delta p_{\mu}:=\bra{\Psi}\mathcal{S}^{\dag}\mathbb{P}_{\mu}(+\infty)\mathcal{S}\ket{\Psi}-\bra{\Psi}\mathbb{P}_{\mu}(-\infty)\ket{\Psi} \ .
\end{equation}

As the redshift is associated with a change in the momentum of the particle, we would expect this to be linked with geodesic motion when no radiation is emitted during the process. In KMOC on a curved background \cite{Adamo:2022rmp}, this is taken into account by adding a completeness relation in (\ref{eq:impulse-redshift}) and restricting this only to massive particle states\footnote{When deriving observables associated with geodesic motion, we should restrict ourselves to tree-level processes only. Thus, it is implicit that the $S$-matrix in (\ref{eq:redshift-2}) is at tree level. While classical, loop contributions in this case are linked to back-reaction effects.}
\begin{equation}\label{eq:redshift-2}
    \Delta p_{\mu}=\int d\Phi(p')\bra{\Psi}\mathcal{S}^{\dag}\ket{p'}p_{\mu}'(+\infty)\bra{p'}\mathcal{S}\ket{\Psi}-\bra{\Psi}\mathbb{P}_{\mu}(-\infty)\ket{\Psi} \ .
\end{equation}
If we then expand the initial state (\ref{eq:initial-state}), we can express the impulse only in terms of on-shell integrals over a $2$-point amplitude \cite{Adamo:2022rmp} 
\begin{equation}
    \Delta p_{\mu}=\int d\Phi(p')\hat{d}^4q \: \hat{\delta}(2 q \cdot p_{0})\bra{p_{0}+q}\mathcal{S}^{\dag}\ket{p'}p_{\mu}'(+\infty)\bra{p'}\mathcal{S}\ket{p_{0}}-p_{0,\mu} \ ,
\end{equation}
where the wavepackets have been removed by taking the classical limit and by sharply localizing the integrand on $p_{\mu}(-\infty)=p_{0,\mu}$. However, as shown in Section~\ref{sec:2pt}, the matrix element of the transfer matrix for a scalar tree-level process in FRW is vanishing in the classical limit. Consequently, the redshift is entirely defined by the background dependence of the momentum operator, and it is not generated by any dynamical scattering process, as it happens instead in plane waves or shock wave backgrounds. Using (\ref{eq:2-point-FRW-vanishing}), it is straightforward to show that the change in momentum is
\begin{equation}
    \Delta p_{\mu}=p_{0, \mu}(+\infty)-p_{0,\mu}=\delta_{\mu 0}\bigg(\sqrt{|\mathbf{p}|^2+m^2a^2_{\infty}}-\sqrt{|\mathbf{p}|^2+m^2}\bigg) \ ,
\end{equation}
in agreement with the geodesic motion. 

It is interesting now to explore what would happen if, instead of truncating the insertion of the completeness relation to massive particle states only, we were to include massless radiative states such as gravitons, associated with the emission of radiation by the point particle while moving on the background \cite{Adamo:2022qci}. 
Let's now look at the structure of the impulse by applying symmetry arguments on amplitudes in FRW. Ignoring quantum processes (e.g. any process involving more than two massive scalars), the completeness relation reads
\begin{align}
    \mathbf{1}= \sum_{n=0}^{\infty}\int d\Phi(p,X_n)  
\ket{p,X_n} \bra{p,X_n}
    \,,
\end{align}
where $X_n$ is the $n$-graviton state and the polarisation sum is implicit. The impulse kernel is then
\begin{align}
\bra{p_2}\mathcal{S}^{\dagger}\mathbb{P}_{\mu}(+\infty)\mathcal{S}\ket{p_1}
=\sum_{n=0}^{\infty}\int d\Phi(p,X_n)   \bra{p_2}\mathcal{S}^{\dagger}\ket{p,X_n} p_{\mu}(+\infty)\bra{p,X_n}\mathcal{S}\ket{p_1}
    \,.
\end{align}
The homogeneity and isotropy of the background imply the factorization of amplitudes ensuring momentum conservation:
\begin{align}
\bra{p',k^{\sigma_1}_1,...,k_n^{\sigma_n}}\mathcal{S}\ket{p}&=2 E_p(-\infty)\: \hat{\delta}^{(3)}\bigg(\mathbf{p}-\mathbf{p}\:'- \mathbf{K}_n \bigg) \: \mathcal{S}_{n+2} \,, 
\end{align}
where $\mathbf{K}_n=\sum^n_{i=1} \mathbf{k}_i$ is the sum of graviton momenta and $\mathcal{S}_{n+2}$ is a scalar quantity defining a generic $(n+2)$-point S-matrix element. Hence, according to unitarity $\mathcal{S}\mathcal{S}^{\dagger}=\mathbf{1}$, we find
\begin{align}
\bra{p_2}\mathcal{S}^{\dagger}\mathbb{P}_{\mu}(+\infty)\mathcal{S}\ket{p_1}
&=\sum_{n=0}^{\infty}\int d\Phi(p,X_n)  \bra{p_2}\mathcal{S}^{\dagger}\ket{p,X_n} (p_{1,\mu}(+\infty)-K_{\mu})\bra{p,X_n}\mathcal{S}\ket{p_1} 
    \nonumber \\
    &= p_{1, \mu}(+\infty)\braket{p_2|p_1} 
    \nonumber \\
    &- \sum_{n=1}^{\infty}\int d\Phi(p,X_n) \bra{p_2}\mathcal{S}^{\dagger}\ket{p,X_n}K^n_{1,\mu}\bra{p,X_n}\mathcal{S}\ket{p_1}
    \,,
\end{align}
in the classical limit where
\begin{align}
    K^n_{1,\mu}=\left( \frac{\mathbf{p}_1\cdot \mathbf{K}_n}{E_1(+\infty)}, -\mathbf{K}_n \right)
    \,.
\end{align}
Thus, the final formula is
\begin{align}
    \Delta p_{\mu}&=p_{0,\mu}(+\infty)-p_{0,\mu}+\Delta p_{\mu}^{\rm rad}\,,
    \label{eq:rad-rec-impulse]}
    \\
    \Delta p_{\mu}^{\rm rad}&=-\sum_{n=1}^{\infty} \int d\Phi(X_n) |\mathcal{S}_{n+2}|^2 K^n_{0,\mu}
    \,.
\end{align}
In addition to the standard redshift, we now have the radiation reaction correction due to graviton emission amplitudes $\mathcal{S}_{n+2}$.

One can notice that radiation reaction correction to the redshift starts already at order $G$,
\begin{equation}
\begin{gathered}
\Delta p_{\mu}^{\rm LO,rad}= -\int d\Phi(k)    
 \sum_{\sigma}\big|\mathcal{S}_{3}^{tree}\big|^2 K^1_{0, \mu} =\mathcal{O}(G)
\end{gathered}
\end{equation}
This contribution is indeed non-vanishing after restoring $\hbar$ factors because the 3-point amplitude on FRW \eqref{3pt_classical} has the same scaling in $\frac{1}{\hbar^{3/2}}$ of a coherent waveshape \cite{Kosower:2018adc}. In computing the waveform associated with this radiation reaction process, we will indeed find it natural to interpret the 3-point amplitude on a cosmological background as the waveshape of a coherent state describing the emitted waveform. Another interesting aspect is that the leading-order correction solely comes from the cut of the one-loop 2-point amplitude. If contributions without the cut were present, they would give rise to a super-classical contribution. However, due to unitarity, such contributions have been already removed in (\ref{eq:rad-rec-impulse]}). This cancellation, ensuring a well-defined classical limit for the impulse, is ubiquitous in the KMOC formalism. In this context, it provides the counterpart, in a cosmological scenario, to the cancellation between box and cross-box diagrams in one-loop calculations for the two-body problem in the post-Newtonian \cite{Neill:2013wsa} and post-Minkowskian approximation \cite{Cheung:2018wkq}.

\subsection{Cosmological waveforms from on-shell data}

As we have seen for the calculation of the redshift, it is crucial to first define a proper Hermitian operator for the observable under consideration, both in the in and out regions. In this case, the field operator for gravitational fluctuations on FRW will have a non-trivial dependence on the conformal time $\eta$. Following \cite{Ford:1977dj}, we introduce a dimensionless waveform operator from the TT mode of the perturbation around the FRW spacetime: 
\begin{equation}\label{eq:quantum-operator-FRW}
 \mathbb{H}_{ij}(x):=\kappa \sum_{\sigma}\int d\Phi(k) \: e^{i \bold{k}\cdot \bold{x}}  a^2(\eta) \Psi(\eta) \hat{a}_{\bold{k},\sigma} \: \varepsilon^{-\sigma}_{ij}(k)+h.c.
\end{equation}
where $\Psi(\eta)$ is the solution to the source free linearized Einstein equations on FRW \eqref{eq:chi-def} with plane wave boundary conditions at $\eta=-\infty$.
Equation (\ref{eq:quantum-operator-FRW}) is defined on the bulk. To define this operator on $\mathcal{I}^{+}$, we need to take the large distance limit of (\ref{eq:quantum-operator-FRW}) while keeping the retarded time $u=\eta-r$ fixed. This is accomplished using a standard stationary-phase
argument as in \cite{Strominger:2017zoo}. Using hatted vectors as unit vectors and writing $\omega=|\mathbf{k}|$ for notational simplicity, we can express (\ref{eq:quantum-operator-FRW}) on $\mathcal{I}^{+}$ as 
\begin{align}\label{eq:wave-quantum-scri} 
    \mathbb{H}_{i j}(r,u,\bold{\hat{x}})|_{\mathcal{I}^{+}}= -\frac{i \kappa}{4\pi r}\sum_{\sigma}\int_{0}^{+\infty} \hat{d}\omega \: (\alpha_{k} e^{-i\omega u}+\beta_{k} e^{+i\omega u}) \hat{a}_{\mathbf{k},\sigma} \varepsilon^{-\sigma}_{i j}(\bold{\hat{x}}) +h.c.\ ,   
\end{align}
where $\mathbf{k}$ should be understood as $\mathbf{k}=\omega \hat{\mathbf{x}}$.
The classical gravitational waves on $\mathcal{I}^{+}$ is then
\begin{equation}\label{eq:waveform-FRW-amplitudes-def}
    h_{ij}(r,u,\bold{\hat{x}})|_{\mathcal{I}^{+}}=\lim_{\hbar \rightarrow 0} \bra{\Psi}S^{\dag}\mathbb{H}_{ij}(r,u,\bold{\hat{x}})|_{\mathcal{I}^{+}}S\ket{\Psi} \ .
\end{equation}
At linear order in the $3$-point amplitude, the classical limit of the mean value of $\hat{a}_{\mathbf{k},\sigma}$ entering in (\ref{eq:waveform-FRW-amplitudes-def}) is rather simple\footnote{We have already performed the on-shell integration against the wavepackets. We should thus consider the momentum $p$ in (\ref{eq:mean-value-a}) as the on-shell momenta of an incoming massive particle.}
\begin{equation}
\begin{gathered}\label{eq:mean-value-a}
\bra{\Psi}\mathcal{S}^{\dag}\hat{a}_{\mathbf{k},\sigma}\mathcal{S}\ket{\Psi}= i\mathcal{A}_{3}^{\sigma}(\omega, \bold{\hat{x}},p)\ ,
\\
    \mathcal{A}_{3}^{\sigma}(\omega, \bold{\hat{x}},p):= -\frac{1}{2}\kappa  \varepsilon^{ij}_{\sigma}(\mathbf{\hat{x}}) [\mathbf{p}\mathbf{p}]^{TT}_{ij}
     \int_{-\infty}^{+\infty} d\eta\, \Psi^*(\eta) \frac{e^{-i \int_{-\infty}^{\eta}d\eta' \frac{\bold{p} \cdot \bar{\bold{k}}}{E_{p}(\eta')}}}{E_{p}(\eta)}  \ . 
    \end{gathered}
\end{equation}
where the massless scalar modes have the incoming boundary condition, consistent with our choice of using in modes for the waveform operator. Equation (\ref{eq:mean-value-a}) can also be interpreted as stating that the final semiclassical state is an eigenstate of the annihilation operator with the $3$-point in the background acting as an eigenvalue. Since coherent states naturally satisfy this property, it is natural to conjecture that the final semi-classical state is
\begin{equation}
\begin{gathered}
    \mathcal{S}\ket{\Psi}=\int d\Phi(p) \phi(p) \ket{p,\alpha}  \ ,
    \\
    \ket{\alpha}=\exp\bigg(-\frac{1}{2}\int d\Phi(\bar{k}) \sum_{\sigma}|\alpha^{\sigma}(\bar{k})|^2 \bigg) \exp\bigg(\int d\Phi(\bar{k})\sum_{\sigma}\alpha^{\sigma}(\bar{k}) \hat{a}_{\bar{\mathbf{k}},\sigma} \bigg)\ket{0_{in}} \ ,
    \end{gathered}
\end{equation}
where the coherent waveshape is given by\footnote{In order to have a well-defined exponent in the classical limit, the waveshape $\alpha^{\sigma}(k)$ needs to scale as $\hbar^{-3/2}$ to compensate for the vanishing measure when expressed in terms of wavenumbers \cite{Kosower:2018adc}. This is consistent with the $\hbar$ scaling of our $3$-point function \eqref{3pt_classical}.}
\begin{equation}
  \alpha^{\sigma}(\bar{k})=i \mathcal{A}^{\sigma}_{3}(\bar{k},p)  \ .
\end{equation}
The same structure can also be understood as an eikonal resummation of $n$-point diagrams with collinear soft gravitons. The existence of this exponentiation in the classical limit is crucial to ensure a well-defined classical limit for the emitted gravitational waves. In fact, a classical gravitational wave is not captured by a single emitted graviton, but by a coherent superposition of gravitons \cite{Cristofoli:2021jas}. 

As for the classical waveform, performing the remaining sum over the polarizations we obtain the final result
\begin{equation}
\begin{gathered}
   h_{ij}(r,u,\bold{\hat{x}})|_{\mathcal{I}^{+}}=  -\frac{\kappa^2 [\mathbf{pp}]^{TT}_{ij} }{8\pi r} \int_{-\infty}^{+\infty} \hat{d\omega}  \: \big(\alpha_k e^{-i \omega u}+\beta_k e^{+i \omega u} \big)  \\ \times \int_{-\infty}^{+\infty} d\eta\, \Psi^*(\eta) \frac{e^{-i \int_{-\infty}^{\eta}d\eta' \frac{\bold{p} \cdot \bar{\bold{k}}}{E_{p}(\eta')}}}{E_{p}(\eta)}  \ .
    \end{gathered}
\end{equation}
We conclude by providing an explicit expression for such quantity for the case of an impulsive FRW background. In this context, the integrand can be read off from (\ref{fin:3-point}) 
\begin{equation}
\begin{gathered}
    h_{ij}(r,u,\bold{\hat{x}})|_{\mathcal{I}^{+}}=  \frac{\kappa^2  [\mathbf{pp}]^{TT}_{ij} }{8\pi r} \int_{-\infty}^{+\infty} \hat{d\omega}  \: i \:\big( \alpha e^{-i \omega u}+\beta e^{+i \omega u}\big) \:  \bigg(\frac{1}{(\omega E_p(-\infty)-\omega \bold{p}\cdot \bold{\hat{x}}-i \varepsilon)} \\
    -\frac{\alpha}{(\omega E_p(+\infty)-\omega \bold{p}\cdot \bold{\hat{x}}+i \varepsilon) }+\frac{\beta}{(\omega E_p(+\infty)+\omega \bold{p}\cdot \bold{\hat{x}}-i \varepsilon) } \bigg)\ ,
    \end{gathered}
\end{equation}
representing an impulsive FRW at $\eta=0$. The frequency integral is singular, and the prescription adopted to deal with the pole at $\omega =0$ is related to a choice of BMS frame, with different prescriptions being related to an ambiguity under BMS supertranslations \cite{DiVecchia:2022owy,Veneziano:2022zwh}. We will use a Feynman prescription using the identity \cite{DiVecchia:2022owy}
\begin{equation}
    \int_{-\infty}^{+\infty} \hat{d\omega} \frac{e^{\pm i \omega u }}{\omega -i \varepsilon}= i \theta(\pm u) \ ,
\end{equation}
This gives the final result\footnote{Strictly speaking, we are covering future null infinity by using Bondi coordinates adapted to the flat case. While the construction of Bondi coordinates for FRW requires delicate work, as shown in \cite{Bonga:2020fhx}, for us, this won't be a significant issue. In fact, we only consider FRW backgrounds of an impulsive type, similar to impulsive plane wave spacetimes. In this case, we can simply avoid the non-asymptotic flatness of a given region by excluding the measurement of radiation in that portion of the Penrose diagram, as done for example for radiative processes on impulsive plane wave spacetimes \cite{Adamo:2022qci}.}
\begin{equation}
\begin{gathered}
    h_{ij}(r,u,\bold{\hat{x}})|_{\mathcal{I}^{+}}= -\frac{4G [\mathbf{pp}]^{TT}_{ij}}{r} \bigg[\theta(-u) \bigg(\frac{\alpha}{E_p(-\infty)-\bold{p}\cdot \bold{\hat{x}}}+\frac{\alpha\beta}{E_p(+\infty)-\bold{p}\cdot \bold{\hat{x}}}+\frac{\alpha\beta}{E_p(+\infty)+\bold{p}\cdot \bold{\hat{x}}} \bigg)\\
    +\theta(u) \bigg( \frac{\beta}{E_p(-\infty)-\bold{p}\cdot \bold{\hat{x}}}+\frac{\alpha^2}{E_p(+\infty)-\bold{p}\cdot \bold{\hat{x}}}+\frac{\beta^2}{E_p(+\infty)+\bold{p}\cdot \bold{\hat{x}}} \bigg) \bigg]
    \end{gathered}
\end{equation}
Our explicit results show that the on-shell 3-point amplitude is indeed non-vanishing and it describes the gravitational wave emission for a particle following geodesic motion in FRW. In Appendix~\ref{sec:classical}, we also compute the wave emission from the purely classical calculation and find the same result. 

\section{Conclusion}

We have derived the cosmological redshift and the waveform emitted by a redshifted point particle using only an S-matrix approach and a notion of classical limit. Classical observables defined on flat spacetime can be derived using only on-shell amplitudes in a rigorous framework known as KMOC \cite{Kosower:2018adc}. However, on a cosmological background, we have noticed that several assumptions in this formalism no longer hold due to the absence of a timelike Killing vector, which motivated us to reformulate this formalism on curved backgrounds \cite{Adamo:2022rmp} more rigorously. 

The absence of a timelike Killing vector yields two features: the non-uniqueness of vacuum and the inequivalence of Hermitian operators in the incoming and outgoing regions of the background. We have discussed that the building blocks of the KMOC formalism are the scattering amplitudes with in/out states being defined by the same definition of the mode functions. We have computed such amplitudes by following the perturbiner approach and correctly reproduced classical dynamics associated with the geodesic motion in the FRW universe. The classical observables receive contributions to the coupling not only from scattering data but also from the time dependence of the Hermitian operators themselves. We should take into account the mismatch of Hermitian operators in the in/out regions to derive the correct formula for the classical observables.


These observations provide a resolution to the conundrum of having a vanishing $2$-point amplitude for a massive scalar on a cosmological background. As argued in \cite{Adamo:2022rmp}, the classical limit of the massive scalar $2$-point amplitude on a background should contain the entire information on the geodesic motion on the associated spacetime. While this statement is definitely true for a shock wave, plane wave, and Schwarzschild background, we have seen that this is not true on a cosmological background. In fact, being careful with the operatorial mismatch between the incoming and outgoing region was crucial to recovering the correct formula for the redshift. It is also interesting to notice that having a $2$-point amplitude vanishing on FRW implies no analogue resummation of flat space-time amplitudes as it is the case instead for shock waves \cite{tHooft:1987vrq}, Schwarzschild \cite{Adamo:2021rfq} and plane waves \cite{Cristofoli:2022phh} backgrounds. We believe that such subtleties, far from being specific to a FRW background, will be relevant for a proper discussion of observables in the self-force expansion from amplitudes \cite{Barack:2023oqp, Adamo:2023cfp, Kosmopoulos:2023bwc, Cheung:2023lnj}.

Another interesting aspect we noticed in our work is that a $3$-point amplitude process, involving a massive scalar emitting a graviton in the background, can be characterized in terms of an on-shell $3$-point process in flat spacetime and a violation of the energy conservation. A distortion of the standard Dirac delta function for the conservation of energy allows for a non-vanishing $3$-point amplitude with real kinematics. This result aligns with the observation made in recent years that $3$-point amplitude processes are related to a rich variety of processes in classical physics. They can be used to generate classical solutions, either by being off-shell in $(1,3)$ signature \cite{Cristofoli:2020hnk} or on-shell in $(2,2)$ signature \cite{Monteiro:2020plf, Monteiro:2021ztt, Sergola:2023mcf}. More recently, by allowing different masses in the external states, they have been shown to capture gravitational wave-absorption phenomena in general relativity \cite{Aoude:2023fdm} and to model the spheroidal harmonic decomposition of solutions to the Teukoslky equations \cite{Chen:2023qzo}. By including large gauge effects in the LSZ, they can be non-vanishing even in Minkowski and provide memory effects \cite{Cristofoli:2022phh} as well as relevant contributions to the one-loop waveform in the post-Minkowskian expansion of the two-body problem \cite{Bini:2024rsy}. Along these lines, it would be interesting to provide an interpretation for our $3$-point process as large gauge effects in the LSZ prescription, as well as exploring whether a BCFW recursion relation is still available when working in a background with no time translation invariance. Yet another direction is to constrain low-energy effective theories for cosmology via the S-matrix, aligned with recent investigations for Lorentz-breaking theories~\cite{Baumann:2015nta, Grall:2021xxm, Aoki:2021ffc, Hui:2023pxc, Creminelli:2023kze}. We leave similar questions for future works.

\acknowledgments

We thank Tim Adamo, Rafael Aoude, Sonja Klisch, Donal O'Connell and Matteo Sergola for useful conversations and comments on the draft. AC is grateful to the Yukawa Institute for Theoretical Physics for their hospitality and for providing a highly stimulating environment where part of this work was developed. The work of KA was supported by JSPS Grants-in-Aid for Scientific Research, No.~20K14468 and No.~24K17046. AC is supported by the Leverhulme Trust (RPG-2020-386). 

\appendix

\section{Classical calculation and equivalence of in-out modes} \label{sec:classical}

It is instructive to compare the on-shell derivation from amplitudes with its purely classical one. In this context, we are looking for a solution to linearized Einstein field equations on FRW sourced by a point particle whose energy-momentum tensor is 
\begin{align}
 T_{\alpha \beta}(y) :=\frac{m}{\sqrt{-g(y)}}\int_{\mathbb{R}} ds \: \delta^{4}(y-x(s)) u_{\alpha}(s) u_{\beta}(s)
 \ .
\end{align}
In the momentum space, the equation of motion is
\begin{align}
\Psi_{ij}''(\eta)+2\frac{a'}{a}\Psi_{ij}^{'}(\eta)+|\bar{\mathbf{k}}|^2 \Psi_{ij}(\eta)=-\frac{\kappa}{2}T^{TT}_{ij}
\,,
\end{align}
where $h_{ij}=a^2\Psi_{ij}$ and $T^{TT}_{ij}(\eta,\bar{\mathbf{k}})$ is the TT part of the energy-momentum tensor. By the use of retarded Green's function\footnote{For more on this Green function, we refer the reader to the detailed analysis in \cite{Chu:2015yua}. We underline also that the often cited Green function in FRW from \cite{Caldwell:1993xw} is not correct. See in particular footnote 7 of \cite{Chu:2015yua}.} on a FRW background, the solution is given by \cite{Jokela:2022rhk}
\begin{align}
    \Psi_{ij}(\eta)&\,=-\frac{\kappa}{2}\int d \eta' G_{\rm ret}(\eta,\eta')T^{TT}_{ij}(\eta')
    \,, \\
    G_{\rm ret}(\eta, \eta')&:=-\theta(\eta-\eta')\frac{a^2(\eta')}{2\bar{\omega} i}(\Psi(\eta)\Psi^*(\eta')-\Psi^*(\eta)\Psi(\eta'))
\end{align}
where $\Psi$ is the same massless mode function we encountered while solving the pertubiner. The Heaviside function can be replaced with unity since we are interested in the wave in the future null infinity. By localizing the integral over the worldline, we can represent the emitted gravitational wave as
\begin{align}
\Psi_{ij}(\eta,\mathbf{x})&=\frac{m\kappa}{2}\int  \frac{\hat{d}^3\bar{\mathbf{k}}}{2i\bar{\omega}} \frac{ds}{a^2(\eta'(s))} e^{i\bar{\mathbf{k}}\cdot(\mathbf{x}-\mathbf{y}(s))} \Big[\Psi(\eta)\Psi^*(\eta'(s))-\Psi^*(\eta)\Psi(\eta'(s))\Big][\mathbf{pp}]_{ij}^{TT} 
\nonumber \\
&=-\frac{i\kappa}{2}\int \frac{\hat{d}^3 \bar{\mathbf{k}}}{2\bar{\omega}} \frac{d\eta'}{E_p(\eta')} e^{i\bar{\mathbf{k}}\cdot(\mathbf{x}-\mathbf{y}(\eta'))} \Big[\Psi(\eta)\Psi^*(\eta'(s))-\Psi^*(\eta)\Psi(\eta'(s))\Big][\mathbf{pp}]_{ij}^{TT} 
\,,
\end{align}
where $\eta \to +\infty$ is understood. Here, we have reparametrized the proper time integral using the conformal time to get the second line and
\begin{align}
    \mathbf{y}(\eta')=\int^{\eta'}_{-\infty} \frac{d\eta''}{E_p(\eta'')} \mathbf{p}
    \,.
\end{align}
Rescaling it by $a^2$ and splitting $[\mathbf{pp}]_{ij}^{TT}$ into the $\pm$ helicity modes, we finally obtain
\begin{align}
    h_{ij}(\eta, \mathbf{x})=\sum_{\sigma}\int d\Phi(\bar{k}) e^{i\bar{\mathbf{k}}\cdot {\mathbf{x}}} a^2(\eta)\Psi(\eta)\epsilon_{ij}^{-\sigma} [i\mathcal{A}^{\sigma}_{\rm 3,cl}] + {\rm c.c.}
\end{align}
with
\begin{align}
\mathcal{A}^{\sigma}_{\rm 3,cl}= -\frac{\kappa}{2}\epsilon_{\sigma}^{ij}[\mathbf{pp}]_{ij}^{TT} \int_{-\infty}^{+\infty} d\eta' \Psi^*(\eta')\frac{e^{-i\int^{\eta'}_{-\infty}d \eta'' \frac{\mathbf{p}\cdot \bar{\mathbf{k}}}{E_p(\eta'')}}}{E_p(\eta')}
\end{align}
and $d\Phi(\bar{k})=\hat{d}^3\bar{\mathbf{k}}/2\bar{\omega}$. Notice that
\begin{align}
    \mathcal{A}^{\sigma}_{\rm 3,cl} =\mathcal{A}^{\sigma}_3
    \,,
\end{align}
in the classical limit.
Therefore, the final answer is in agreement with the calculation derived from purely on-shell methods on the background. 

As we have discussed in Sec.~\ref{sec:KMOC}, the in/out vacua do not contain the classical wave so either in vacuum or out vacuum can be used to prepare the semiclassical initial state in the strict classical limit. Let us reintroduce the in/out labels. Our computations so far are all based on the in-vacuum and the replacement ``in'' with ``out'' gives the computation based on the out-vacuum in the KMOC. Let's refer to the classical calculations to see that the final result remains the same under this replacement. In classical computation, the mode functions appear in the retarded Green's function. One can easily see
\begin{align}\label{Green_in/out}
    G_{\rm ret}(\eta, \eta')&=-\theta(\eta-\eta')\frac{a^2(\eta')}{2\bar{\omega} i}(\Psi_{in}(\eta)\Psi_{in}^*(\eta')-\Psi_{in}^*(\eta)\Psi_{in}(\eta'))
    \nonumber \\
    &=-\theta(\eta-\eta')\frac{a^2(\eta')}{2\bar{\omega} i}(\Psi_{out}(\eta)\Psi_{out}^*(\eta')-\Psi_{out}^*(\eta)\Psi_{out}(\eta'))
\end{align}
where the in/out mode functions are related by the Bogoliubov transformation. Since the amplitude-based computation agrees with the classical computation, the relation \eqref{Green_in/out} ensures that the final result does not change under the replacement of in/out-vacuum in the classical limit.

\bibliographystyle{JHEP}
\bibliography{biblio}

\end{document}